    \documentclass[preprint,showpacs,preprintnumbers,superscriptaddress,amsmath,amssymb]{revtex4}

    \usepackage[dvips]{graphicx}
    \usepackage{mathrsfs}
    \usepackage{amssymb}
    \usepackage{dcolumn}
    \usepackage{bm}
    \usepackage{subfigure}
    \usepackage{floatflt}
    \usepackage{float}
    \usepackage{rotating}
    \usepackage{multirow}
    \usepackage[bookmarksnumbered,bookmarksopen,colorlinks,citecolor=blue,linkcolor=blue]{hyperref} %
\begin{document}

    \preprint{00-000}

    \title{Sensitive dependence of isotope and isobar distribution of limiting temperatures on symmetry energy}

    \author{Li Ou}
    \email{only.ouli@gmail.com}
    \affiliation{College of Physics and Technology, Guangxi Normal University, Guilin, 541004, People's Republic of China}

    \author{Min Liu}%
    \affiliation{College of Physics and Technology, Guangxi Normal University, Guilin, 541004, People's Republic of China}

    \author{Zhuxia Li}%
    \affiliation{China Institute of Atomic Energy, Beijing, 102413, P. R. China}

    \date{\today}

    \begin{abstract}

    The mass, isotope, and isobar distributions of limiting
    temperatures for finite nuclei are investigated by using a thermodynamics approach
    together with the Skyrme energy density functional.
    The relationship between the width of the isotope (isobar) distribution of limiting temperatures
    and the stiffness of the density dependence of the symmetry energy clearly is observed.
    The nuclear symmetry energy with smaller slope parameter $L_{\rm{sym}}$
    causes a wider the isotope (isobar) distribution of limiting temperatures.
    The widths of the isotope (isobar) distributions of limiting
    temperatures could be useful observables for exploring the information of the density dependence
    of the nuclear symmetry energy at finite temperatures.

    \end{abstract}

    \maketitle

    The nuclear symmetry energy plays a crucial role for understanding nuclear phenomena
    and for exploring the equation of state (EOS) for isospin asymmetric nuclear matter.
    Significant efforts have been devoted to constrain the symmetry energy
    at both high densities\cite{Danielewicz02,Fuchs06}
    and subsaturation densities \cite{Tsang04,Liutx07,Famiano06,Tsang09,Danielewicz2009,Walker2010}.
    Up to now, some constraints on symmetry energy at subnormal densities have already been obtained
    from different experimental measurements that include nuclear structure and reactions
    \cite{Klimkiewicz07,Centelles09,Tsang09,Baldo04,LiuM10,Tsang12}.
    However, the uncertainties of the density dependence of nuclear symmetry energy are still large.
    More information on the nuclear symmetry energy is still required
    for understanding the structures of nuclei far away from the $\beta$-stability line,
    heavy-ion collisions, supernova explosions, and neutron star properties \cite{Liba08,Steiner05}.

    The energy per nucleon in uniform nuclear matter can be written as
    $E(\rho,\delta)=E_0(\rho,\delta=0)+E_{\rm{sym}}(\rho)\delta^2$,
    where $\delta=(\rho_{n}-\rho_{p})/\rho$, $\rho_{n}$, $\rho_{p}$, and
    $\rho$ are the neutron, proton, and nucleon densities, respectively.
    $E_{\rm{sym}}(\rho)$ describes the density dependence of symmetry energy
    and can be expended as
    \begin{align}
    E_{\rm{sym}}(\rho)=E(\rho_{0})+\textstyle{L_{\rm{sym}} \over 3}\left(\textstyle{\rho-\rho_{0} \over \rho_{0}}\right) 
    +\textstyle{K_{\rm{sym}} \over 18}\left(\textstyle{\rho-\rho_{0} \over \rho_{0}}\right)^{2}+\cdots,
    \end{align}
    where $L_{\rm{sym}}=3\rho_{0}\textstyle{\partial E_{\rm{sym}}(\rho) \over \partial\rho}|_{\rho_{0}}$ and
    $K_{\rm{sym}}=9\rho_{0}^2\textstyle{\partial^2 E_{\rm{sym}}(\rho) \over \partial^2\rho}|_{\rho_{0}}$
    denote the slope and curvature parameters, respectively.
    On the other hand, the symmetry energy also depends on the temperature,
    which is also of fundamental importance for the liquid-gas phase transition of asymmetric nuclear matter,
    the dynamical evolution mechanisms of massive stars and the supernova explosion.
    The behavior of the temperature dependence of the symmetry energy is less well
    understood \cite{XuJ07, LiBA06, Kowalski07, Moustakidis07, OuL11},
    which compares the symmetry energy at zero temperature.

    It is found that the calculated limiting temperature sensitively depends
    on the stiffness of the EOS (the incompressibility),
    critical temperature, surface tension, {\it{et al}}. \cite{Natowitz95,Song91,Kelic06, Lizx04}.
    From experimental observations of limiting temperature, Natowitz {\it{et al}}.
    successfully derived the critical temperature and the
    incompressibility of isospin symmetric nuclear
    matter \cite{Natowitz02,Natowitz02l}.
    Further, Li and Liu\cite{Lizx04} pointed out that the isotope
    distribution of limiting temperatures sensitively depended on the isospin
    dependent part of interaction. But the effects of the
    isoscalar and isovector parts of the EOS on limiting temperature are entangled in their paper.
    In this Rapid Communication,
    to manifest the isospin effect, we first investigate the limiting
    temperatures of nuclei in isotope and isobar chains.  Then we investigate the
    correlation between the isotope (isobar) distribution of the limiting
    temperatures and the density dependence of symmetry energy.
    In addition, we attempt to extract the information of the density
    dependence of symmetry energy at finite temperatures from available experimental data.

    We use the same model as that used in Refs. \cite{Lizx04, Lium03,Jaqaman89}.
    Within this model a hot nucleus is considered as
    a spherical liquid droplet of uniformly distributed nuclear matter at constant temperature.
    This liquid droplet is in thermal equilibrium with the surrounding vapor.
    The thermal, mechanical, and chemical equilibriums between the liquid droplet and the surrounding vapor
    are required, which leads to a set of standard coexistence equations,
    \begin{align}\label{phaseeq}
    \mu_p(T,\rho_L,\delta_L)&=\mu_p(T,\rho_V,\delta_V),\\
    \mu_n(T,\rho_L,\delta_L)&=\mu_n(T,\rho_V,\delta_V),\\
    P(T,\rho_L,\delta_L)&=P(T,\rho_V,\delta_V).
    \end{align}
    The subscript $L$ refers to the liquid droplet, and the subscript $V$ refers to the
    surrounding vapor. For simplification, the Coulomb interaction
    is screened in the calculation of the pressure and the
    proton chemical potential of the surrounding vapor.
    The maximum temperature at which the coexistence equations have
    solutions is the limiting temperature.
    The chemical potential of the nucleon of species $q$
    can be written as
    \begin{align}\label{chem1}
    \mu_{q}(T,\rho,\delta)=&u_{q}(T,\rho,\delta)+T\sum_{n=1}^{\infty}
    \textstyle{n+1 \over n}b_{n}(1 \pm \delta)^n \left(\textstyle{\lambda_{T}^{3} \over g_{s,I}} \rho
    \right)^{n}
     \nonumber \\
    &+T\ln(1 \pm \delta)
    +T\ln\textstyle{\lambda_{T}^{3} \over g_{s,I}}\rho+\varepsilon_{\rm{Coul}}\delta_{q,p},
    \end{align}
    where the symbol ``$+$'' stands for neutrons and the symbol ``$-$'' stands for protons.
    The $\lambda_{T}$ is the
    effective thermal wavelength of the nucleon, which reads
    \begin{eqnarray}\label{lambdaT}
    \lambda_T  = \left( {\textstyle{{2\pi \hbar ^2 } \over {m_{q}^* T}}}\right)^{1/2}.
    \end{eqnarray}
    $m^{*}_{q}$ and $u_{q}$ are the effective mass and the
    single-particle potential energy, respectively,
    $b_n$'s are the coefficients of the virial series for the ideal Fermi gas,
    $g_{s,I}=4$ is the spin-isospin degeneracy,
    and $\varepsilon_{\rm{Coul}}$ is the Coulomb energy term.
    The total pressure of droplet is written as
    \begin{eqnarray}
    P(T,\rho,\delta)=P_{\rm{bulk}}+P_{\rm{Coul}}+P_{\rm{surf}}.
    \end{eqnarray}
    The bulk pressure of the nucleus can be calculated by \cite{SJLee01}
    \begin{align}
    P_{\rm{bulk}}&=\sum_{q}\left[\left(\textstyle{5 \over 3}\textstyle{1 \over 2 m^*_{q}}
    - \textstyle{1 \over 2m_{q}}\right)\hbar^2\tau_{q} +{\textstyle{1 \over 2}} u_{q} \rho(1\pm \delta) \right] -U,
   \end{align}
    where $U$ and $\tau_q$ in $P_{\rm{bulk}}$ are the potential density
    and the kinetic-energy density of the nucleon, respectively,
    $P_{\rm{Coul}}$ is the pressure contributed by the Coulomb
    interaction, and $P_{\rm{surf}}$ is the pressure contributed by
    the surface tension that includes a symmetry-surface term suggested in Refs.
    \cite{Chabanat,Myers} (which is called Surf2 in Ref.
    \cite{Lizx04}). The critical temperature for infinite nuclear
    matter is taken as 17 MeV, referenced from Ref. \cite{Natowitz02l} where
    $T_c=16.6\pm 0.86$ MeV.

\begin{figure}[h]
        \centering
            \includegraphics[width=8.5cm]{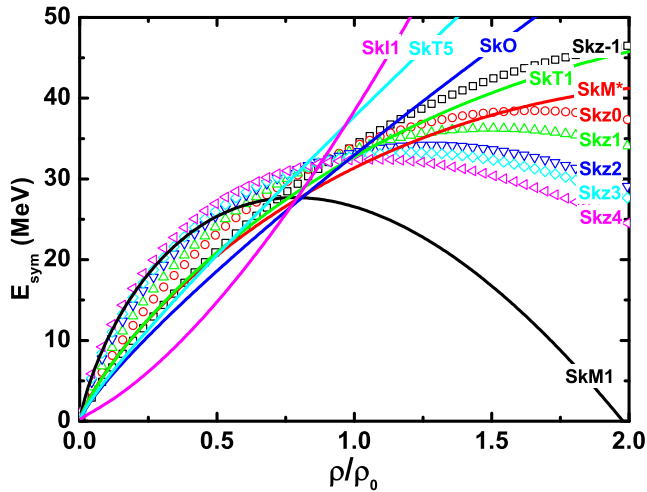}
            \caption{(Color online) Density dependence of symmetry energy
            with various Skyrme interactions.}
            \label{fig1}
        \end{figure}

    The effective Skyrme interaction is adopted in this Rapid Communication,
    and the expressions of $m^*_{q}$, $u_{q}$, and $\tau_{q}$ can be found in Ref. \cite{Lizx04}.
    To study  the effect of symmetry energy on limiting temperature,
    29 sets of Skyrme interactions are selected in the calculations
    with the values of incompressibility $K_{\infty}=230\pm30$~MeV
    and quite different values of $L_{\rm{sym}}$ and $K_{\rm{sym}}$.
    In Table \ref{table1}, we list the slope parameter $L_{\rm{sym}}$, asymmetry coefficient $a_{\rm{s}}$
    at temperatures of $T$=0/5 MeV, curvature parameter $K_{\rm{sym}}$, and
    incompressibility module $K_{\infty}$ at temperature of $T$=0 MeV,
    predicted by these Skyrme interactions.
    The Skyrme interactions are sorted by the ascending order with slope parameter $L_{\rm{sym}}$ at zero temperature.
    \begin{table}
    \renewcommand\arraystretch{0.9}
    \caption{\label{table1}
    Slope parameter $L_{\rm{sym}}$, asymmetry coefficient $a_{\rm{s}}$ at temperatures of $T$=0/5 MeV,
    curvature parameter $K_{\rm{sym}}$, and incompressibility module $K_{\infty}$ at temperature of $T$=0 MeV,
    predicted by different Skyrme interactions.}
    \begin{ruledtabular}
    \begin{tabular}{ccccc}
    Version   &  $L_{\rm{sym}}$  &  $a_{\rm{s}}$&  $K_{\rm{sym}}$& $K_{\infty}$  \\ \hline
    SkM1\cite{SkM1}    &$-$31.17/$-$29.95   & 26.48/26.22    & $-$383  & 216  \\
    SVII\cite{SVII}    &   $-$9.28/$-$7.89&  27.86/27.46 &  $-$488&   367  \\
    Skz4\cite{Skz}   &  4.89/6.17      & 32.36/31.78    & $-$246  & 230 \\
    Skz3\cite{Skz}    & 14.19/15.37     & 32.80/32.31    & $-$243  & 230   \\
    Skz2\cite{Skz}    & 20.50/21.59     & 33.31/32.96    & $-$256  & 230  \\
    Skz1\cite{Skz}    & 33.06/34.17     & 33.66/33.46    & $-$235  & 230   \\
    BSk9\cite{BSk9}    & 40.24/41.37 &  30.79/30.22 &  $-$148&   231  \\
    Skz0\cite{Skz}    & 42.56/44.02     & 34.09/34.14    & $-$231  & 230   \\
    SLy7\cite{Chabanat}    &   47.72/48.92 &  32.91/32.41 &  $-$116&   230  \\
    SkM$^*$\cite{SkMs} & 50.13/51.34     & 31.46/31.20    & $-$151  & 216 \\
    SkT3\cite{SkT}    &   56.77/57.73 &  32.20/31.62 &  $-$134&   236 \\
    SkT2\cite{SkT}    &   57.58/58.54 &  32.70/32.12 &  $-$136&   236 \\
    SkT1\cite{SkT}    & 57.60/58.56     & 32.72/32.14    & $-$136  & 236 \\
    KDE0v1\cite{KDE}  &   58.92/60.11 &  35.55/35.04 &  $-$130&   232 \\
    SKRA\cite{SKRA}   &   59.96/58.09 &  32.72/32.45 &  $-$133&   214 \\
    SQMC650\cite{SQMC} &   59.65/60.89 &  35.04/34.82 &  $-$168&   222 \\
    SV-sym32\cite{SVSYM} &   61.09/62.61 &  33.62/33.41 &  $-$144&   233 \\
    Skz-1\cite{Skz}   & 62.40/64.16     & 34.27/34.44    & $-$171  & 230 \\
    NRAPR\cite{Steiner05}   &   62.45/63.53 &  33.40/33.11 &  $-$117&   222 \\
    LNS\cite{LNS} &   62.65/63.75  & 34.41/34.07 &  $-$127&   214 \\
    SQMC700\cite{SQMC} &   63.67/64.81 &  34.20/33.92 &  $-$133&   214 \\
    MSL0\cite{MSL0}    &   64.21/65.36 &  31.55/31.21 &  $-$97&    233 \\
    Ska35s20\cite{Dutra}&   65.06/66.00 &  34.59/34.01 &  $-$122&   240 \\
    Ska25s20\cite{Dutra}&   66.58/67.56 &  34.93/34.35 &  $-$120&   221 \\
    Skxs20\cite{Skx} &   72.55/73.59 &  37.27/36.74 &  $-$123&   207 \\
    SkO\cite{SkO}     &  81.70/82.79     & 32.95/32.47    & $-$43   & 224 \\
    SkT5\cite{SkT}    & 100.11/101.09   & 37.10/36.52    & $-$26   & 202 \\
    SkI5\cite{SkI}    &   128.01/128.70 &   36.11/35.60 &  156&    256\\
    SkI1\cite{SkI}    & 160.74/161.91   & 38.24/37.73    & 234   & 244
    \end{tabular}
    \end{ruledtabular}
    \end{table}
    Figure \ref{fig1} presents the density dependence of the symmetry energy with some Skyrme interactions,
    which describe the possible behavior of the symmetry energy predicted by different theories.
    Especially, we select the Skz-series because these interactions
    have almost the same isoscalar part but varied isovector part in the EOS,
     which are especially useful for studying the symmetry energy effect.

\begin{figure}[h]
        \centering
            \includegraphics[width=8.5cm]{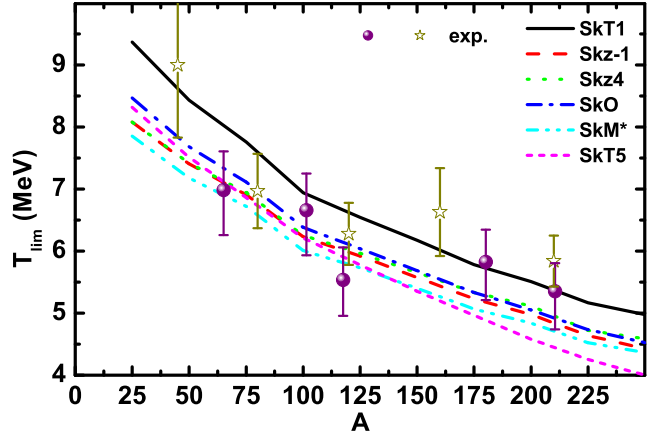}
            \caption{(Color online) Mass distributions of the limiting
            temperatures for $\beta$-stability nuclei calculated with different Skyrme interactions.}
            \label{fig2}
        \end{figure}

    Figure \ref{fig2} shows the mass distributions of limiting
    temperatures of nuclei along the $\beta$-stability line with
    $Z=0.5A-3\times10^{-3} A^{5/3}$, calculated with different Skyrme interactions.
    The data, which are extracted from a number of different experimental measurements
    and only for symmetric or slightly asymmetric nuclei,  are taken from Refs. \cite{Natowitz02,Natowitz02l}.
    From Fig. \ref{fig2}, one can see the influence of $K_{\infty}$ on the limiting temperature.
    The calculation with the stiffer EOS obtains the higher limiting
    temperature, which is consistent with other investigations\cite{Song91,Kelic06, Lizx04}.
    As expected, the behavior of the symmetry energy does not significantly influence
    the mass dependence of the limiting temperatures for the $\beta$-stability nuclei.
    For example, the calculation results with Skz-1 and Skz4 are almost the same,
    although the corresponding isovector parts are quite different.
    To reveal the isospin energy effect more clearly, we further
    study the limiting temperatures of nuclei in isotope and isobar chains.

\begin{figure} [h]
        \centering
            \includegraphics[width=8.5cm]{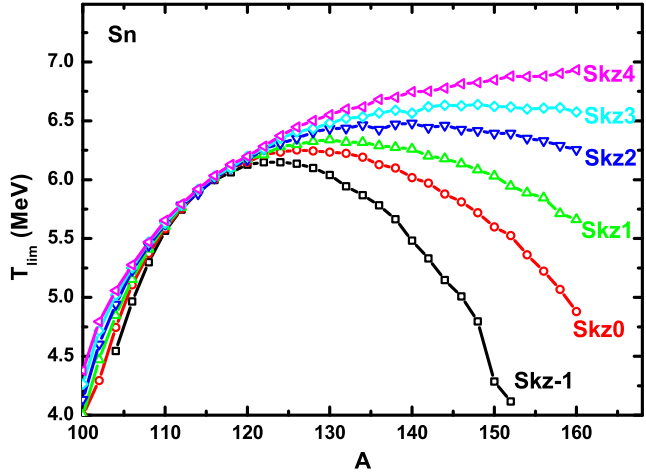}
            \caption{(Color online) Isotope distributions of the limiting temperatures for Sn
            calculated  with Skz-series Skyrme interactions.}
            \label{fig3}
        \end{figure}

    Figure \ref{fig3} shows the isotope distributions of limiting
    temperatures for Sn isotopes calculated with Skz series Skyrme interactions.
    From the figure, one can see that all isotope distributions of
    limiting temperatures appear to be inverted parabolas.
    On the left side of the parabolas, the limiting temperatures of the nuclei
    increase with the neutron numbers since the Coulomb potential is reduced.
     On the right side, the limiting temperature decreases
    with the extra richness of the neutrons because the nuclei become unstable
    due to symmetry energy that is  too strong.
    The competition between the Coulomb energy and the symmetry energy
    leads to the parabolic shape of the isotope distribution.
    All the curves intercross around $^{116}$Sn which is the
    corresponding $\beta$-stability isotope of Sn. This is due to the fact that
    the parameters of each Skz interaction are fitted to the
    properties of nuclei near the $\beta$-stability line.
    Furthermore, all Skz interactions have the same isoscalar part in the EOS,
    which leads to the similar behaviors for the $\beta$-stability nuclei.
    For the isotopes far away from the $\beta$-stability line, the difference between the
    limiting temperatures calculated with different Skz interactions
    becomes large. The most clear and interesting feature
    shown in Fig. \ref{fig3} is as follows: The softer the symmetry energy is,
    the broader the distribution of the limiting temperature is,  and we get a
    higher limiting temperature for the neutron-rich isotope.

\begin{figure} [h]
        \centering
            \includegraphics[width=8.5cm]{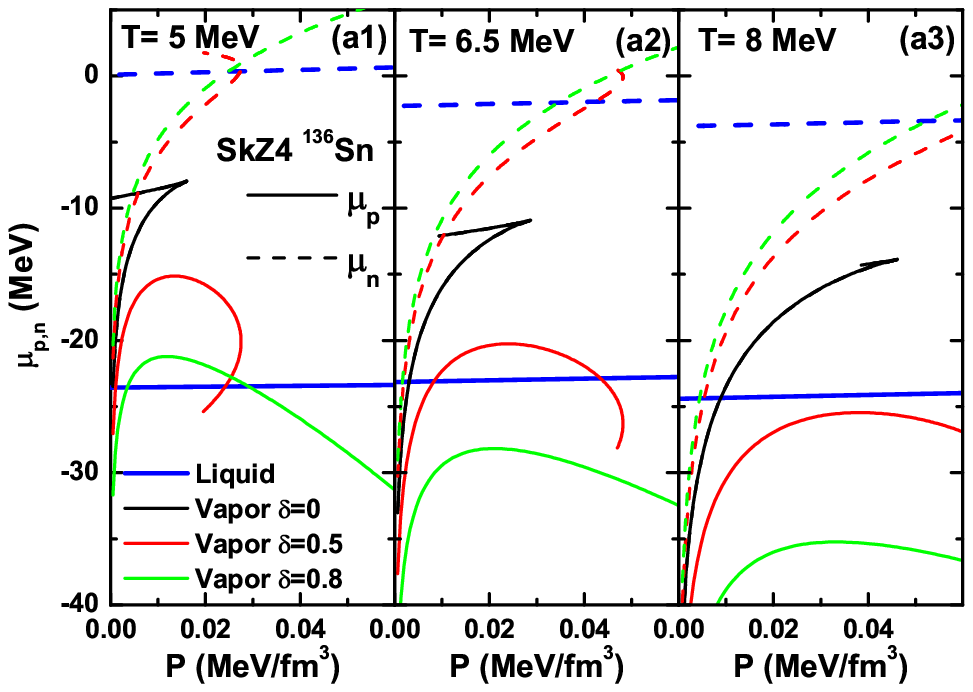} \\
            \includegraphics[width=8.5cm]{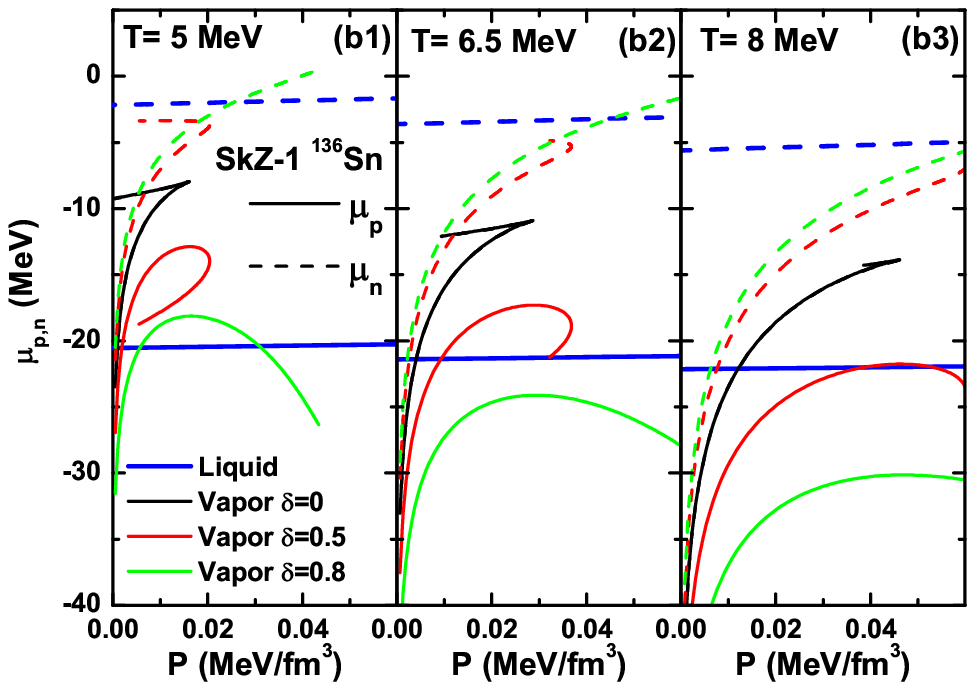}
            \caption{(Color online) Correlations of $\mu_{n}\sim P$ and $\mu_{p}\sim P$ for $^{136}$Sn
    nucleus and surrounding vapor at $T$=5, 6.5, and 8 MeV, calculated
    with Skz4 (upper panel) and Skz-1 (bottom panel).}
            \label{fig4}
        \end{figure}

    To understand the effect of symmetry energy on the
    limiting temperature, in Fig. \ref{fig4}, we present
    the correlations of $\mu_{n}\sim P$ and $\mu_{p}\sim P$ for $^{136}$Sn
    and the surrounding vapor at $T$=5, 6.5 and 8 MeV calculated
    with Skz4 and Skz-1, respectively.
    Since the isospin asymmetry $\delta_{V}$ is not fixed, we take three different values
    $\delta_{V}$=0.0, 0.5, and 0.8. One can find
    that the proton and neutron chemical potentials for the vapor (which
    is low density and neutron rich) decrease with temperature.
    The neutron chemical potential of the vapor is higher, and the proton chemical potential of
    the vapor is lower with Skz4 than the corresponding results by using Skz1 because the
    symmetry energy for Skz4 is much softer than that for Skz-1. If the solution for
    coexistence equations exists, there simultaneously should be the intersects
    between the liquid and the vapor curves for both
    $\mu_{n}\sim P$ and $\mu_{p}\sim P$. Because of the effect
    of the symmetry energy, one sees that in Figs.\ref{fig4}(a1)
    and \ref{fig4}(b1) at $T$ = 5 MeV, there simultaneously exist intersects
    for $\mu_{n}\sim P$ and $\mu_{p}\sim P$ with $\delta_{V}$=0.5 and 0.8
    for the Skz4 case but only at $\delta_{V}$=0.8 for the Skz-1 case.
    In Figs.\ref{fig4}(a2) and \ref{fig4}(b2) at $T$=6.5 MeV,
    there simultaneously exist intersects for $\mu_{n}\sim P$ and $\mu_{p}\sim P$
    at $\delta_{V}$=0.5, and intersects do not exist for $\delta_{V}$=0.8
    with Skz4 because the proton chemical potential
    with $\delta_{V}$ =0.8 is reduced too much. For the Skz-1 case,
    there simultaneously is no intersect for the $\mu_{n}\sim P$ and $\mu_{p}\sim P$ curves.
    At $T$ = 8 MeV there simultaneously is no intersect between
    the liquid and the vapor curves for $\mu_{n}\sim P$ and $\mu_{p}\sim P$
    because the vapor $\mu_{p}\sim P$
    curve becomes too low to cross over the liquid $\mu_{p}\sim P$
    curve for the Skz4 case and, for the Skz-1 case, the vapor $\mu_{n}\sim P$ curve
    becomes too low to cross over the liquid $\mu_{n}\sim P$ curve.
    From the above discussions, it can be understood that the softer
    symmetry energy increases the $\mu_{n}$ of the vapor, which makes it
    possible for the vapor to be in equilibrium with the liquid at a higher temperature.
    Thus, the higher limiting temperature is obtained for the softer
    symmetry energy case.

    The experiment S254, conducted at the SIS heavy-ion synchrotron at GSI Darmstadt,
    was devoted to study the isotope effects in projectile fragmentation at relativistic energy\cite{Sfienti09}.
    The collisions of 600 MeV/nucleon $^{124}$Sn, $^{107}$Sn and $^{124}$La on $^{\rm{nat}}$Sn
    were performed, the limiting temperatures for nuclei with the same
    $A/Z$ but $Z_{\rm{bound}}/Z_{\rm{proj}}$ intervals [0.6,0.8] for
    $^{124}$Sn and [0.55,0.75] for the neutron-poor cases ($^{107}$Sn
    and and $^{124}$La) were extracted.
    According to this experimental measurement, the spectator systems are most
    likely populated in the bin of nuclei with the same $A/Z$ but only $75\%$
    of the projectile mass\cite{Sfienti09,Pochodzalla95}.
    Thus, we investigate the limiting temperatures for the isotope chain of $_{38}$Sr
    and isobar chain of $^{93}A$ by attempting to
    obtain some information on the density dependence of the symmetry energy
    at finite temperatures with these data.

\begin{figure} [h]
        \centering
            \includegraphics[width=8.5cm]{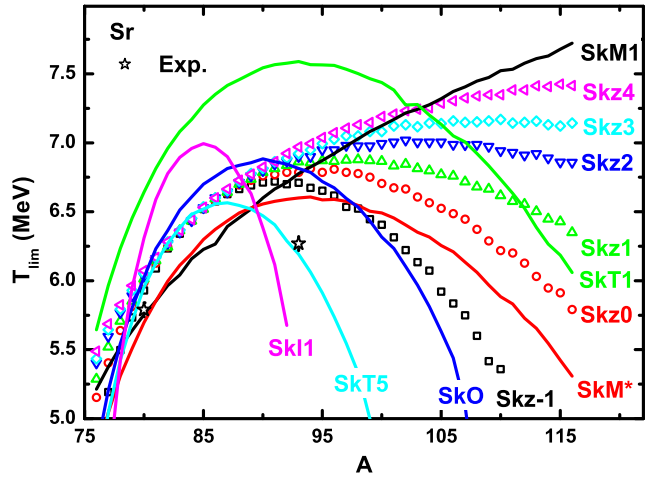}
            \caption{(Color online) Isotope distributions of the
        limiting temperatures for Sr calculated with various Skyrme interactions.}
            \label{fig5}
        \end{figure}

\begin{figure} [h]
        \centering
            \includegraphics[width=8.5cm]{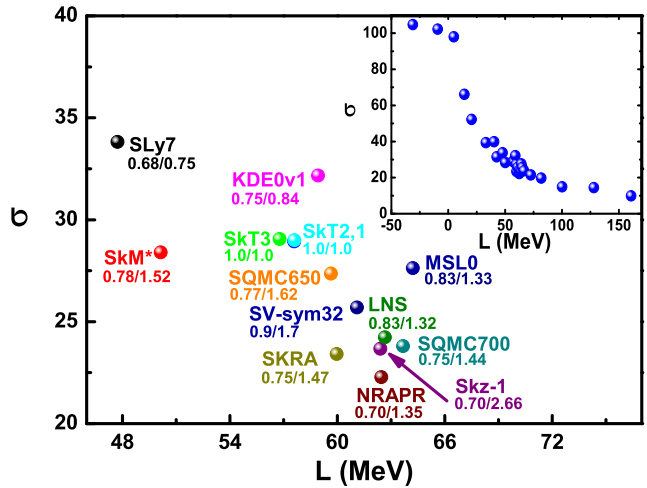}
            \caption{(Color online) Correlation between distribution width $\sigma$ and $L_{\rm{sym}}$ of the symmetry energy,
            see the text for details.}
            \label{fig6}
        \end{figure}

    Figure \ref{fig5} presents the isotope distributions of
    limiting temperatures for Sr calculated with various Skyrme interactions.
    The data are taken from \cite{Sfienti09}.
    One can see that all Skz family interactions fail to reproduce
    the experimental data for $^{80,93}$Sr. The results
    are overestimated. It seems that the symmetry energy is too soft.
    To describe the experimental data,
    more interactions with various stiffnesses of symmetry energy,
    which includes those sugguested by Dutra {\it{et al}}.\cite{Dutra},
    are included in the calculations.
    It seems that the results with SkT5 reproduce the data reasonably well.
    However the calculation results look a little messy, even if just
    the partial results are shown in the figure. We believe that this
    chaos is caused mainly by different isoscalar parts of the EOS
    as shown in Fig.~\ref{fig2}.
    It is known that both the isovector and the isoscalar part influence the results.
    To reduce the influence from the isoscalar part of the interaction,
    we only concentrate on the
    shapes of the isotope distributions of limiting temperatures
    rather than their absolute values.
    To quantitatively describe the shape of the distribution, we introduce the width of
    distribution $\sigma$. $\sigma$ is obtained by fitting the isotope distribution
    of limiting temperatures with a three-parameter Gaussian function,
    \begin{eqnarray}
    g(A)=\textstyle{a \over \sqrt{2\pi}\sigma}\exp\left[\textstyle{-(A-A_c)^2 \over 2\sigma^2} \right].
    \end{eqnarray}
    The correlation between distribution width $\sigma$ and $L_{\rm{sym}}$ of the symmetry energy
    is illustrated in the inner figure in Fig. \ref{fig6}.
    One can see that a softer symmetry energy obtains a
    wider distribution of the limiting temperature, which is independent on the
    isoscalar part of the EOS.
    We also note that there are some fluctuations for width $\sigma$ within the range
    of $L_{\rm{sym}}$ from 48 to 65 MeV.
    To explore the reason for the fluctuation, we show the results in
    the enlarged image. For each calculated point,
    we present the label of the Skyrme interaction, the effective
    mass ($m^*/m$) and the effective mass splitting (EMS) ratio
    ($m^*_n/m^*_p$) at the saturation density. It seems that the
    fluctuations have a certain relation with the EMS. From the figure,
    we find that, for the Skyrme interactions with similar $L_{\rm{sym}}$ values,
    the Skyrme interactions with $m^*_n<m^*_p$ (SLy7, KDE0v1)
    obtain larger $\sigma$, and those with $m^*_n>m^*_p$ obtain smaller $\sigma$'s.
    It is actually understandable as the kinetic energy also
    contributes to the chemical potential and pressure of the nuclei
    in which the effective mass of the proton (neutron) is involved.

\begin{figure}
        \centering
            \includegraphics[width=8.5cm]{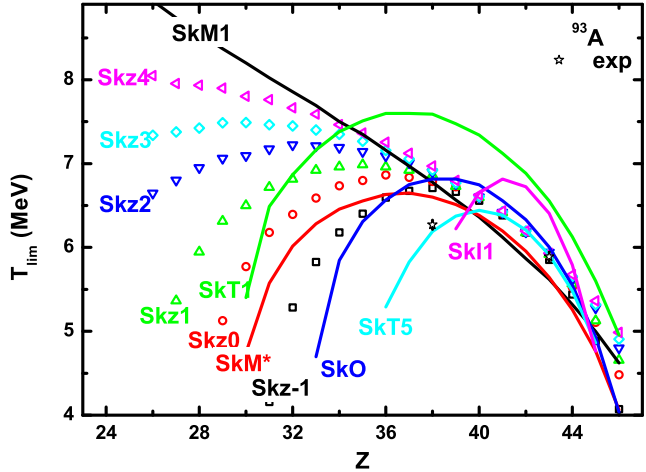}
            \caption{(Color online) Isobar distributions of the
        limiting temperatures for $^{93}A$ calculated with various Skyrme interactions.}
            \label{fig7}
        \end{figure}

    We perform the same calculations for the isobar chain of $^{93}A$.
    From the results for the $^{93}A$ isobar shown in Fig.~\ref{fig7},
    the consistent conclusion can be obtained as that for the Sn isotopes, i.e.,
    the softer symmetry energy obtains a wider distribution of limiting
    temperature.

    These investigations indicate that the widths of the isotope and
    isobar distributions of limiting temperatures are closely
    correlated to the density dependence of the symmetry energy at finite temperatures.
    The neutron-proton EMS also has certain influences on
    the widths of the isotope and isobar distributions of limiting temperatures,
    which can also provide us with information for the neutron-proton EMS.
    From Figs. \ref{fig5} and \ref{fig7}, we note that the curves calculated with SkT5
    roughly pass through the data points for $^{80}$Sr, $^{93}$Sr and $^{93}$Tc and a
    better agreement is obtained compared with other interactions.
    As $L_{\rm{sym}}$ and $a_{\rm{s}}$ are correlated, some
    investigations produce a range of acceptable values (see Ref. \cite{Tsang12} for a recent summary).
    The values of $L_{\rm{sym}}$ and $a_{\rm{s}}$ for SkT5 are located at the large side of the acceptable values.
    Moreover, see Fig. \ref{fig1}, the SkT5 has the symmetry energy almost linearly depending on
    density (i.e., the small $K_{\rm{sym}}$, which is even less
    constrained up to now). However, as we know that the limiting
    temperature depends on both the isoscalar and the isovector parts,
    the symmetry energy can not be constrained uniquely by two data points
    in the isotope and isobar distributions of the limiting temperatures.
    To obtain the experimental information for the width of the
    isotope (isobar) distribution of limiting temperatures, at least
    three points are needed. Thus, at least one more datum
    is required to determine the width of distribution in addition
    to $^{80,93}$Sr or $^{93}$Sr and $^{93}$Tc. For this purpose,
    $^{83}$Sr-$^{86}$Sr or $^{93}_{40}$A-$^{93}_{42}$A should be the
    best candidates.

    To summarize, the mass, isotope, and isobar distributions of limiting
    temperatures are investigated by using 29 sets of Skyrme interactions.
    The correlation between the width of the isotope (isobar) distribution
    of limiting temperatures and the slope parameter $L_{\rm{sym}}$ of
    the symmetry energy clearly is observed from the calculations.
    A softer symmetry energy causes a wider isotope (isobar)
    distribution of limiting temperatures. The neutron-proton
    EMS also slightly influences the width of the distribution.
    As a helpful observable, the width of the isotope(isobar)
    distribution of limiting temperatures should be measured
    for obtaining the information of the isovector part of the EOS, not only
    the momentum-independent part, but also the momentum-dependent part.
    With concerning for the available experimental data of the isotope Sr and isobar
    $^{93}A$ chain, at least one more datum point
    is required to determine the width of distribution. For this purpose
    $^{83}$Sr-$^{86}$Sr or $^{93}_{40}A$-$^{93}_{42}A$ should be the
    best candidates in addition to $^{80,93}$Sr or $^{93}$Sr and $^{93}$Tc.

    We thank Professor N. Wang for a careful reading of the manuscript.
    This work was supported by the National Natural Science Foundation
    of China (under Grants No. 11005022,
    No. 11365004, 
    No. 11365005, 
    No. 11075215, 
    No. 11005003, 
    No. 11275052, 
    No. 11375062),  
    was supported partly by MOST of China (973 Program with Grant No. 2013CB834404)
    and by the Doctor Startup Foundation of Guangxi Normal University.


\begin{thebibliography}{99}
    \bibitem{Danielewicz02} P. Danielewicz, R. Lacey, and W. G. Lynch, Science {\bf298}, 1592 (2002).

    \bibitem{Fuchs06} C. Fuchs, Prog. Part. Nucl. Phys {\bf56}, 1 (2006).

    \bibitem{Tsang04} M. B. Tsang,  {\it{et al}}.,
      Phys. Rev. Lett. {\bf92}, 062701 (2004).

    \bibitem{Liutx07} T. X. Liu,  {\it{et al}}.,
    Phys. Rev. C. {\bf76}, 034603 (2007).

    \bibitem{Famiano06} M. A. Famiano,  {\it{et al}}.,
    Phys. Rev. Lett. {\bf97}, 052701 (2006).

    \bibitem{Tsang09}M. B. Tsang,
     Y. Zhang, P. Danielewicz, M. Famiano, Z. Li, W. G. Lynch and A. W. Steiner,
    Phys. Rev. Lett. {\bf 102}, 122701 (2009).

    \bibitem{Danielewicz2009} P. Danielewicz and J. Lee Nucl. Phys.  {\bf A818}, 36 (2009).

    \bibitem{Walker2010} P. M. Walker, S. Lalkovski, and P. D. Stevenson, Phys. Rev. C {\bf 81}, 041304 (2010).

    \bibitem{Tsang12} M. B. Tsang,  J. R. Stone, F. Camera, {\it{et al}}., Phys. Rev. C {\bf 86}, 015803 (2012).

    \bibitem{Klimkiewicz07} A. Klimkiewicz, N. Paar, and P. Adrich  {\it{et al}}., Phys. Rev. C {\bf 76},
    051603 (2007).

    \bibitem{Centelles09} M. Centelles, X. Roca-Maza, X. Vi\~{n}as, and M. Warda, Phys.
    Rev. Lett. {\bf 102}, 122502 (2009).

    \bibitem{Baldo04} M. Baldo, C. Maieron, P. Schuck, and X. Vi\~{n}as, Nucl. Phys.  {\bf A736}, 241 (2004).

    \bibitem{LiuM10} M. Liu, N. Wang, Z.-X. Li, and F.-S. Zhang, Phy. Rev. C {\bf 82}, 064306 (2010).

    \bibitem{Liba08} B.-A. Li, L.-W. Chen, and C. M. Ko, Phys. Rep. {\bf464}, 113 (2008).

    \bibitem{Steiner05} A. W. Steiner, M. Prakash, J.M. Lattimer, and P. J. Ellis, Phys. Rep. {\bf 411}, 325 (2005).

    \bibitem{XuJ07} J. Xu, L.-W. Chen, B.-A. Li, and H.-R. Ma, Phys. Rev. C {\bf 75} (2007) 014607;
                    Phys. Rev. C {\bf 77} (2008) 014302.

    \bibitem{LiBA06} B.-A. Li, and L. Chen, Phys. Rev. C {\bf 74} (2006) 034610.

    \bibitem{Kowalski07} S. Kowalski, J. B. Natowitz, S. Shlomo, {\it{et al}}.,
    Phys. Rev. C {\bf 75} (2007) 014601.

    \bibitem{Moustakidis07} C. C. Moustakidis, Phys. Rev. C {\bf 76} (2007) 025805.

    \bibitem{OuL11} L. Ou, Z. Li, Y. Zhang, and M. Liu, Phys. Lett. B {\bf 697}, 246 (2011).

    \bibitem{Natowitz95} J. B. Natowitz, K. Hagel, R. Wada, Z. Majka, P. Gonthier, J. Li,
    N. Mdeiwayeh, B. Xiao, and Y. Zhao,  Phys. Rev. C {\bf 52}, R2322 (1995).

    \bibitem{Song91} H. Q. Song and R. K. Su, Phys. Rev. C {\bf 44}, 2505 (1991).

    \bibitem{Kelic06} A. Keli\`{c}, J.B. Natowitz, and K.-H. Schmidt, Eur. Phys. J.
    A {\bf 30}, 203 (2006).

    \bibitem{Lizx04} Z. Li, and M. Liu, Phys. Rev. C {\bf69}, 034615 (2004).

    \bibitem{Natowitz02} J. B. Natowitz,
     R. Wada, K. Hagel, T. Keutgen, M. Murray, A. Makeev, L. Qin, P. Smith, and C. Hamilton,
     Phys. Rev. C {\bf 65}, 034618 (2002).

    \bibitem{Natowitz02l} J. B. Natowitz, K. Hagel, Y. Ma, M. Murray, L. Qin, R. Wada, and J. Wang,
    Phys. Rev. Lett. {\bf 89}, 212701 (2002).

    \bibitem{Lium03} M. Liu, Z. Li, and J. Liu, Chin. Phys. Lett. {\bf20}, 1076 (2003).

    \bibitem{Jaqaman89} H. R. Jaqaman, Phys. Rev. C {\bf39}, 169 (1989); {\bf40}, 1677 (1989).

    \bibitem{SJLee01} S. J. Lee, and A. Z. Mekjian, Phys. Rev. C {\bf63}, 044605 (2001).

    \bibitem{Chabanat} E. Chabanat, P. Bonche, P. Haensel, J. Meyer, and R.Schaeffer,
    Nucle. Phys. {\bf A627}, 710 (1997); {\bf A635}, 231 (1998).

    \bibitem{Myers} W. D. Myers, and W. J. Swiatecki, Nucl. Phys. {\bf 81}, 1 (1966).

    \bibitem{SkM1} J. M. G. G\'{o}mez and M. Casas, Few-Body Syst., Suppl. {\bf 8}, 374 (1995);
                    X. Li, and P.-H. Heenen, Phys. Rev. C {\bf 54}, 1617 (1996).

    \bibitem{SVII} M. J. Giannoni and P. Quentin, Phys. Rev. C {\bf 21}, 2076 (1980).

    \bibitem{Skz} J. Margueron, J. Navarro, and N. V. Giai, Phys. Rev. C {\bf 66}, 014303 (2002).

    \bibitem{BSk9} S. Goriely, M. Samyn, J. M. Pearson, andM. Onsi, Nucl. Phys.  {\bf A750}, 425 (2005).


    \bibitem{SkMs} J. Bartel, P. Quentin, M. Brack, C. Guet, and H.-B. Hakansson, Nucl. Phys. {\bf A386}, 79 (1982).

    \bibitem{SkT} F. Tondeur, M. Brack, M. Farine, and J. M. Pearson, Nucl. Phys.  {\bf A420}, 297 (1984).

    \bibitem{KDE} B. K. Agrawal, S. Shlomo, and V. K. Au, Phys. Rev. C {\bf 72}, 014310 (2005).

    \bibitem{SKRA} J. M. G. G\'{o}mez, C. Prieto, and J. Navarro, Nucl. Phys.  {\bf A549}, 125 (1992).

    \bibitem{SQMC} P. A. M. Guichon, H. H. Matevosyan, N. Sandulescu, and A. W. Thomas, Nucl. Phys.  {\bf A772}, 1 (2006).

    \bibitem{SVSYM} P. Kl\"{u}pfel, P.-G. Reinhard, T. J. B\"{u}rvenich, and J. A. Maruhn, Phys. Rev. C {\bf 79}, 034310 (2009).


    \bibitem{LNS} L. G. Cao, U. Lombardo, C. W. Shen, and N. Van Giai, Phys. Rev. C {\bf 73}, 014313 (2006).

    \bibitem{MSL0} L.-W. Chen, C. M. Ko, B.-A. Li, and J. Xu, Phys. Rev. C {\bf 82}, 024321 (2010).

    \bibitem{Dutra} M. Dutra, O. Louren\c{c}o, J. S. S\'{a}Martins, and A. Delfino,  Phys. Rev. C {\bf 85}, 035201 (2012).

    \bibitem{Skx} B. A. Brown, G. Shen, G. C. Hillhouse, J. Meng, and A. Trzci\'{n}ska, Phys. Rev. C {\bf 76}, 034305 (2007).

    \bibitem{SkO} P.-G. Reinhard, {\it{et al}}.,
    Phys. Rev. C {\bf 60}, 014316 (1999).

    \bibitem{SkI} P.-G. Reinhard and H. Flocard, Nucl. Phys.  {\bf A584}, 467 (1995).

    \bibitem{Sfienti09} ALADIN2000 Collaboration, C. Sfienti {\it{et al}}.,
     Phys. Rev. Lett. {\bf 102}, 152701 (2009).

    \bibitem{Pochodzalla95} J. Pochodzalla, {\it{et al}}.,
    Phys. Rev. Lett. {\bf 75}, 1040 (1995).


    \end{thebibliography}
    \end{document}